\documentclass[english,aps,pra,reprint,noshowpacs,superscriptaddress,nofootinbib]{revtex4-1}   
\usepackage[T1]{fontenc}	
\usepackage[latin9]{inputenc}	
\usepackage{geometry}		
\usepackage{graphicx}
\usepackage[above,below]{placeins}	
\usepackage{times}
\usepackage{amsmath,amsthm}

\usepackage{hyperref}
\usepackage{array}
\hypersetup{colorlinks=true,urlcolor=blue,citecolor=blue,linkcolor=blue}   
\begin{document}

\title{Methods to Simplify Object Tracking in Video Data}
\author{Orban, C. M.}\email{orban@physics.osu.edu}
\affiliation{Physics Department, The Ohio State University, 191 W Woodruff Ave, Columbus, OH 43210}
\affiliation{Physics Department, The Ohio State University at Marion, 1461 Mount Vernon Ave, Marion, OH 43302}
\author{Zimmerman, S.}
\affiliation{Department of Mathematics, The Ohio State University, 231 West 18th Avenue, Columbus, OH 43210}
\affiliation{Department of Mathematics, The Ohio State University at Marion, 1461 Mount Vernon Ave, Marion, OH 43302}
\author{Kulp, J. T.}
\affiliation{Physics Department, The Ohio State University, 191 W Woodruff Ave, Columbus, OH 43210}
\author{Boughton, J.}
\affiliation{Great Oaks Career Campuses, 110 Great Oaks Drive, Cincinnati, OH, 45241}
\author{Perrico, Z.}
\affiliation{University of Mount Union, Alliance, OH, 44601}
\author{Rapp, B.}
\affiliation{University of Mount Union, Alliance, OH, 44601}
\author{Teeling-Smith, R.}
\affiliation{University of Mount Union, Alliance, OH, 44601}

\maketitle

\section{Introduction}

\label{sec_intro}  Recent years have seen an explosion of interest in analyzing the motion of objects in video data as a way for students to connect the concepts of physics to something tangible like a video recording of an experiment \cite{pivotinteractives,TrackerVideo}. Inexpensive devices can now record videos at 120 or 240 frames per second. And among instructors there is an appetite for substantially re-thinking the objectives of introductory labs (e.g. \cite{Etkina2015,Ansell_Selen2016,Holmes_Wieman2018}).


Generally, the goal of a student activity involving analysis of video data is to obtain the $(x,y)$ position of a particular object in as many frames of the video as possible. Once obtained, this data can be used to infer velocities, acceleration and any number of other quantities like momentum or energy. A variety of software exists for students to look at individual frames and click on the object to infer the $(x,y)$ position (e.g. \cite{verniervideo,TrackerVideo,NewtonDV,PlaygroundPhysics,jstrack}). For longer videos or videos recorded at a high frame rate this task can become tedious. Some, but not all, of these tools include a capability to automatically identify the position of the object in the frame \cite{verniervideo,TrackerVideo,NewtonDV}. But it is not unusual, especially when inexperienced users are recording the video and configuring the program, for these algorithms to struggle to ``lock on" to the moving object. In this paper, we both include some general advice to help object tracking algorithms locate an object 
and we provide our own algorithms that are simpler and potentially more effective than the sophisticated image processing algorithms that are currently being used. Our algorithms are built into a free and open source program called the STEMcoding Object Tracker (available at \url{http://go.osu.edu/objecttracker}) which works through the browser and is compatible with a variety of operating systems including Chromebooks. An interesting ``feature" of this tool is that it does \emph{not} automatically calculate the velocity and acceleration of the object being tracked. Instead students must build their data analysis skills to extract these quantities from the raw $(x,y)$ versus time data.


\section{The Challenge of Object Tracking}

\begin{figure*}
\includegraphics[width=6in]{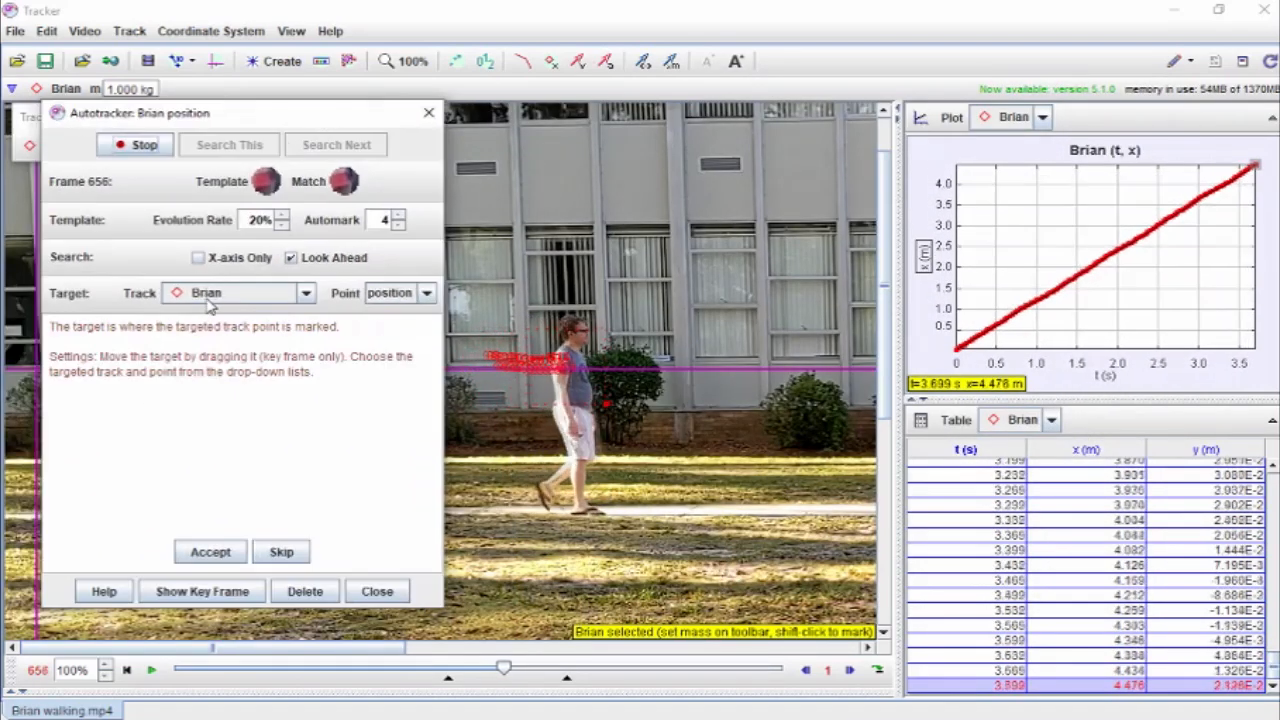}
\caption{A screenshot from the program Tracker Video where a red piece of paper on the shoulder of a walking person is being tracked across the screen. The "Autotracker" window on the left shows the template image being searched for in each frame and the matching image found in a given frame.} \label{fig:brian}
\end{figure*}

Fig.~\ref{fig:brian} shows a screenshot from a video tutorial on object tracking from the Let's Code Physics YouTube channel by Dr. Brian Lane \cite{BrianWalking}. Dr. Lane is using a program called Tracker Video to analyze a video where he is walking across the sidewalk \cite{TrackerVideo}. 
In the analyzed video, Dr. Lane has a red piece of paper taped to his right shoulder. This piece of paper is the object being tracked. 

According to the Tracker Video help guide\cite{TrackerHelp}, ``Autotracker works by creating one or more template images of a feature of interest and then searching each frame for the best match to that template." In the video tutorial, Dr. Lane explains that it is important to define the template image of the object not at the center of the paper but rather at one of the corners so that the object tracking algorithm searches for a splotch of red next to some gray, which is the color of his shirt. With this hint, the object tracking works well and the plot of $x$ versus time shows that Dr. Lane is walking at an approximately constant velocity as expected. 

Although this is just a brief moment in the tutorial video it underscores the difficulty of what automatic object tracking is attempting to do. The background of Fig.~\ref{fig:brian}, for example, has many different features that the program potentially needs to scan through as it searches for the the red piece of paper. Further complicating matters, the paper itself will look different in each frame due to changes in lighting and shadows. In our own experimentation with object tracking using a different program -- specifically the Vernier Video Physics app -- we found it to be surprisingly difficult to get the program to ``lock on" to a moving object. 


We provide some general thoughts on object tracking and then we describe two different algorithms that can be used to track an object moving in front of a solid colored background. We demonstrate this method on a difficult object tracking problem -- that of a roll of tape that is spinning as it moves through the air. We implement these algorithms in a program called the STEMcoding Object Tracker which is freely available at \url{http://go.osu.edu/objecttracker}.



\section{Simplifying the Task}

In this paper we advocate for three methods to simplify the challenge of automatically tracking objects:

\begin{enumerate}
    \item Use a solid colored background, like a black poster board, a green screen, or a white classroom wall.
    \item Use an object that is bright red, green, or blue.
    \item Use an object tracking algorithm that searches for specific colors instead of searching for a template image.
\end{enumerate}

In the present work we show two examples of videos recorded with a solid colored background -- a basketball being tossed in front of a black curtain and a blue roll of tape moving in front of a white background. In our tests we found that object trackers could easily find the objects in each frame.



The reason we advocate for a black or white background with a bright red, green or blue object in the foreground is that video data is stored in an array of pixels, and each pixel stores three values -- one for red, one for green and one for blue. For historical reasons\footnote{An 8-bit unsigned integer can take any integer value between 0 and 255}, each color has a value between 0 and 255. So pure red would be $(255,0,0)$, pure green would be $(0,255,0)$ and pure blue would be $(0,0,255)$. In this scheme, pure white is $(255,255,255)$ and pure black is $(0,0,0)$. 

From an information perspective the largest contrast that one can achieve would be a bright red, green or blue object in front of a dark background. One can also use a solid colored blue, green, red or white background so long as the object is a different color. In our own tests we found that using a green screen background with a basketball in the foreground is another good combination. This is because green screens (which are also called ``chromakey") use a shade of green that is close to $(0,255,0)$. 

Even if using a template image instead of a reference color for the object, the contrast between the object color and the background color helps the object tracking algorithm sift through all the regions of the image to find the best match.

\section{Simplifying the Algorithm}

The STEMcoding Object Tracker includes two different object tracking algorithms both of which focus on the color of the object as compared to that of the background. The user is prompted to click the object at three different locations in a frame, at which time the program stores three separate values for each color variables $r$, $g$, and $b$ corresponding (ideally) to the object color. The user is then instructed to click the background at three different locations in the same frame and three more values for each of $r$, $g$, and $b$ corresponding to the background color are collected. The values of $r$, $g$, and $b$ are each averaged individually for the object and the background. One can instead manually define the average color variables for the background and the object if desired. 

In our program are two different ways of taking these average colors and searching through the pixels to locate the object: (1) The Color Distance Method, and (2) Brightness Agnostic Mode.

\subsection{The Color Distance Method}

As physics instructors, it is convenient to think about the rgb colors like a three element vector
\begin{equation}
    \vec{c} = r \, \hat{i} + g \, \hat{j} + b \,  \hat{k}
\end{equation}
where r, g and b are integers between 0 and 255.  Each pixel has some color $\vec{c}$ which can be compared to the average color of the object ($\vec{c}_{\rm obj}$) and the background ($\vec{c}_{\rm bg}$). Here, $r_{\rm obj}$, $g_{\rm obj}$, and $b_{\rm obj}$ are the average color values of the object as discussed above, and $r_{\rm bg}$, $g_{\rm bg}$, and $b_{\rm bg}$ are the averages for the background.

A simple way to compare the vectors is by taking the norm of the difference between them -- in other words calculate the color distance, for example,
\begin{equation}
    |\vec{c} - \vec{c}_{\rm obj}| = \sqrt{(r - r_{\rm obj})^2 + (g - g_{\rm obj})^2 + (b - b_{\rm obj})^2}
\end{equation}
and
\begin{equation}
    |\vec{c} - \vec{c}_{\rm bg}| = \sqrt{(r - r_{\rm bg})^2 + (g - g_{\rm bg})^2 + (b - b_{\rm bg})^2}.
\end{equation}
The ``color distance method'' flags a particular pixel as object if $|\vec{c} - \vec{c}_{\rm obj}|$ is smaller than $|\vec{c} - \vec{c}_{\rm bg}|$.

While straightforward, the color distance method can fail if the object is dimly lit or not particularly bright colored. Gray-ish pixels on a solid black background can have rgb values similar to $(100,100,100)$ which may be closer to the object color than to a pure black background of $(0,0,0)$. In general, gray pixels will have $r \sim g \sim b$ so any amount of gray in the background can potentially be confused for the object.

\subsection{Brightness Agnostic Method}

The problem with the color distance method is essentially that the brightness of the pixel affects whether the pixel is considered object or part of the background, especially if one is using a dark or bright white background. One approach to avoid this would involve normalizing the color vectors by their magnitude (in other words dividing each color channel by the brightness). This operation converts the color vectors to unit vectors (i.e. vectors of magnitude 1). Such color vectors are independent of the brightness of the pixel. However, following this scheme would be problematic for pure black pixels $(0,0,0)$ because the normalization would involve dividing by zero.

Instead we propose a scheme where a constant $255/2$ is subtracted from each component of the color vectors so that the color channels take non-integer values between $-255/2$ and $+255/2$ as follows. Writing
\begin{equation}
    \vec{c}_0 = (255/2)\hat{i} - (255/2)\hat{j} - (255/2)\hat{k},
\end{equation}
we have the following:
\begin{equation}
     \vec{c} - \vec{c}_0 = (r - 255/2)\hat{i} + (g - 255/2) \hat{j} + (b - 255/2)\hat{k}.
\end{equation}
This vector can be normalized by its magnitude to create a unit vector $\hat{c}$
\begin{equation}
    \hat{c} = \frac{\vec{c} - \vec{c}_0}{|\vec{c} - \vec{c}_0|}.
\end{equation}
%
%
By defining $\hat{c}$ in this way we can be sure that $|\hat{c}|$ evaluates to 1, including for black $(0,0,0)$ pixels. Before computing $\hat{c}$ for every pixel in the frame, the algorithm will compute $\hat{c}_{\rm obj}$ from the object colors and $\hat{c}_{\rm bg}$ from the background colors as follows
\begin{equation}
    \hat{c}_{\rm obj} = \frac{\vec{c}_{\rm obj} - \vec{c}_0}{|\vec{c}_{\rm obj} - \vec{c}_0|}
    \qquad
    \hat{c}_{\rm bg} = \frac{\vec{c}_{\rm bg} - \vec{c}_0}{|\vec{c}_{\rm bg} - \vec{c}_0|}.
\end{equation}
The $\hat{c}$ from each pixel can then be compared to the object and the background using a dot product.
\begin{equation}
      \hat{c} \cdot \hat{c}_{\rm obj} =  |\hat{c}| |\hat{c}_{\rm obj}| \cos \theta_{\rm obj} =  \cos \theta_{\rm obj}   
\end{equation}
\begin{equation}
     \hat{c} \cdot \hat{c}_{\rm bg} =  |\hat{c}| |\hat{c}_{\rm bg}| \cos \theta_{\rm bg} =  \cos \theta_{\rm bg}.
\end{equation}
If the pixel color and object color are exactly the same, then $\hat{c}$ and $\hat{c}_{\rm obj}$ are the same vector, and so $\cos \theta_{obj} = 1$. If instead the pixel color is very different than the object color, then $\cos \theta_{obj}$ should be close to $-1$. Similarly, if the pixel color and the background color are exactly the same, $\cos \theta_{\rm bg} = 1$, and if they are very different, $\cos \theta_{bg}$ should be close to $-1$.

Our algorithm flags pixels as being contained in the object if $\cos \theta_{\rm obj} > \cos \theta_{\rm bg}$. 
This inequality implies that the angle between the pixel's normalized color vector ($\hat{c}$) and the object's normalized color vector ($\hat{c}_{\rm obj}$) is smaller than the the angle between $\hat{c}$ and $\hat{c}_{\rm bg}$.

\begin{figure*}
\includegraphics[width=5.5in]{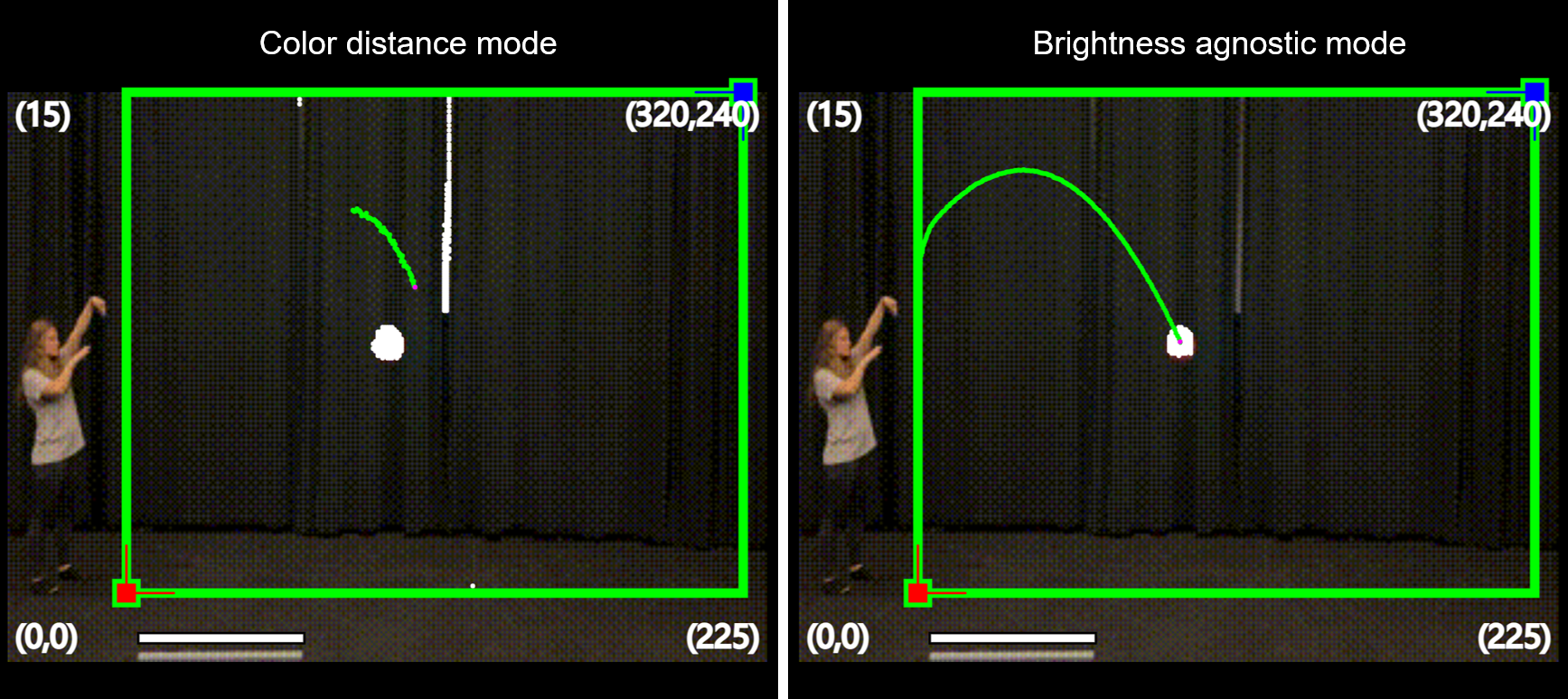} 
\caption{A basketball moving in front of a black curtain is being tracked using the two different modes of the STEMcoding Object Tracker. Pixels identified as object are highlighted in white. The center of all object pixels is shown with a purple dot and the movement of that center is shown with a green line. The green box defines the region of interest. Using color distance mode (left panel), a line of gray pixels in the background is incorrectly flagged as object. Using brightness agnostic mode (right panel), only the orange-red pixels of the basketball are identified as object.}\label{fig:bball}
\end{figure*}

Fig.~\ref{fig:bball} shows the result of using this algorithm with a basketball moving in front of a black curtain. Although there is a thin line of gray in the frame that the color distance method often flags as part of the object, this mistake is avoided in brightness agnostic mode because the brightness of a pixel does not impact whether it is included as part of the object of the background. Note that, if we digitally remove the thin line of gray, the color distance method can track the basketball as well as the brightness agnostic method can.


\subsection{Comparing to Tracker Video}

As mentioned earlier, the object finding method used in Tracker Video \cite{TrackerVideo} uses an image of the object rather than an rgb color. Of course the image of the object contains the rgb colors so certainly Tracker Video is using colors in the frame as part of its algorithm to find the object. Therefore in practice the method used in Tracker Video is similar to what we propose here. An essential difference is that, once the object is located in Tracker Video, the program uses the image of the object in the most recent frame as the new "template image" for the following frame. If at some frame the program fails to locate the object correctly, it may be difficult for it to find it again in later frames. However, in the methods we propose, the ability of the program to locate the object in one frame does not affect its ability to locate the object later on as the only data used are the colors of the object and the background, and these colors are kept constant for the duration of the video.


\section{Tracking Pixels on a Rotating Hollow Object}

\begin{figure*}
    \centering
    \includegraphics[width=4in]{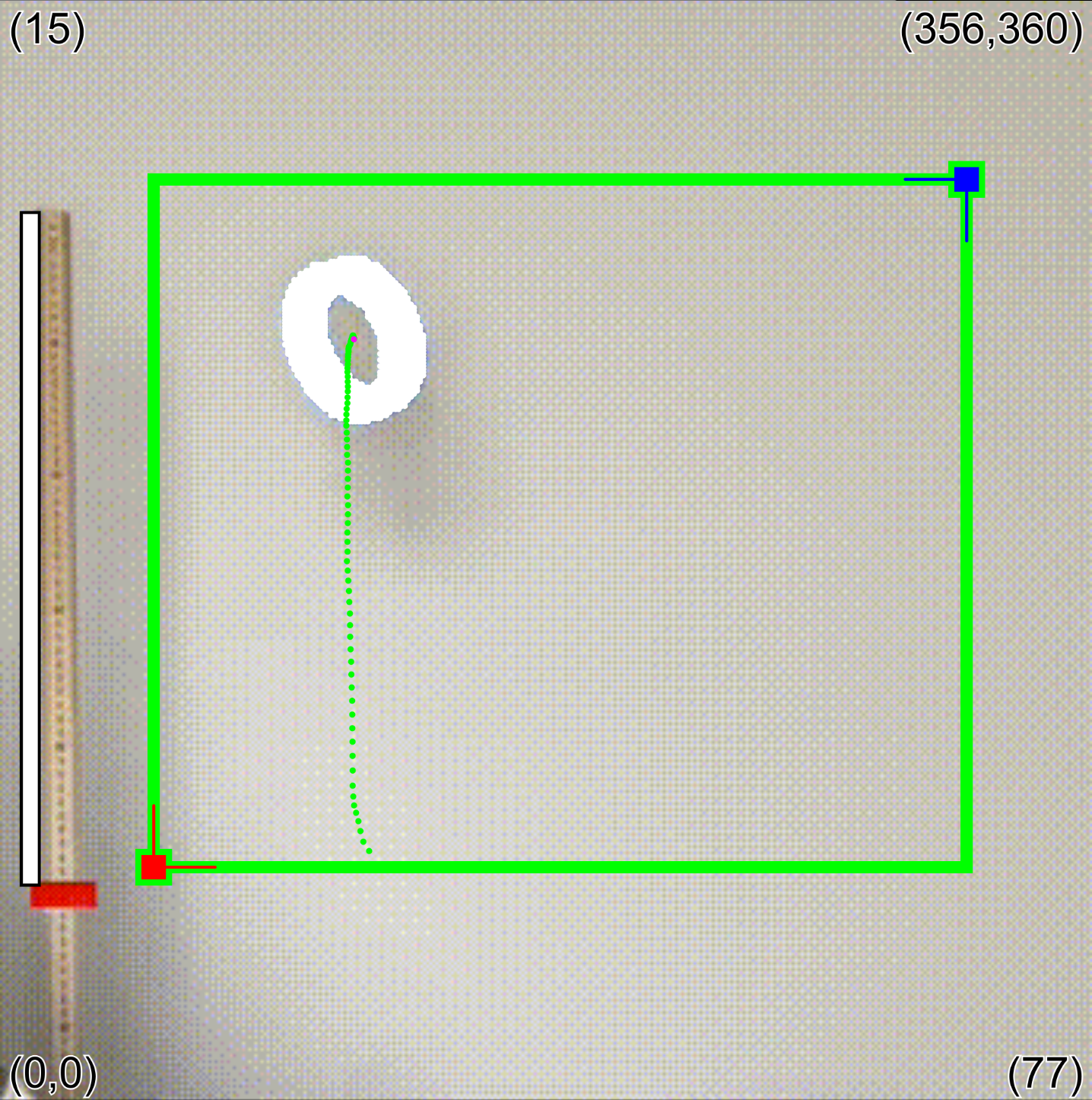}
    \caption{A still frame from a blue roll of painters tape being tossed in the air. Each blue-colored pixel is identified as object and a pure white pixel is drawn over that pixel. The center of all of these object pixels is shown with a purple dot which is a useful way to follow the motion of the center of mass. The green box shows the region of interest.}
    \label{fig:tape}
\end{figure*}

A potential advantage of tracking colors is that the object does not need to be spherical like a ball or flat and facing the camera like the red paper in the example highlighted in Fig.~\ref{fig:brian}. Fig.~\ref{fig:tape} shows an example where a roll of blue-colored painters tape has been tossed straight upwards while it rotates about its axis. 

Both the color distance and brightness agnostic methods can easily find the blue pixels on a white-ish background and, in Fig.~\ref{fig:tape}, each of these blue pixels are covered with a pure white pixel to indicate that the program has identified these pixels as part of the object. Because the object is circular, the center of these pixels happens to be the center of mass of the object. As is well known, the center of mass of the object should behave like a simple projectile. 

\begin{figure}
    \centering
    \includegraphics[width=3.35in]{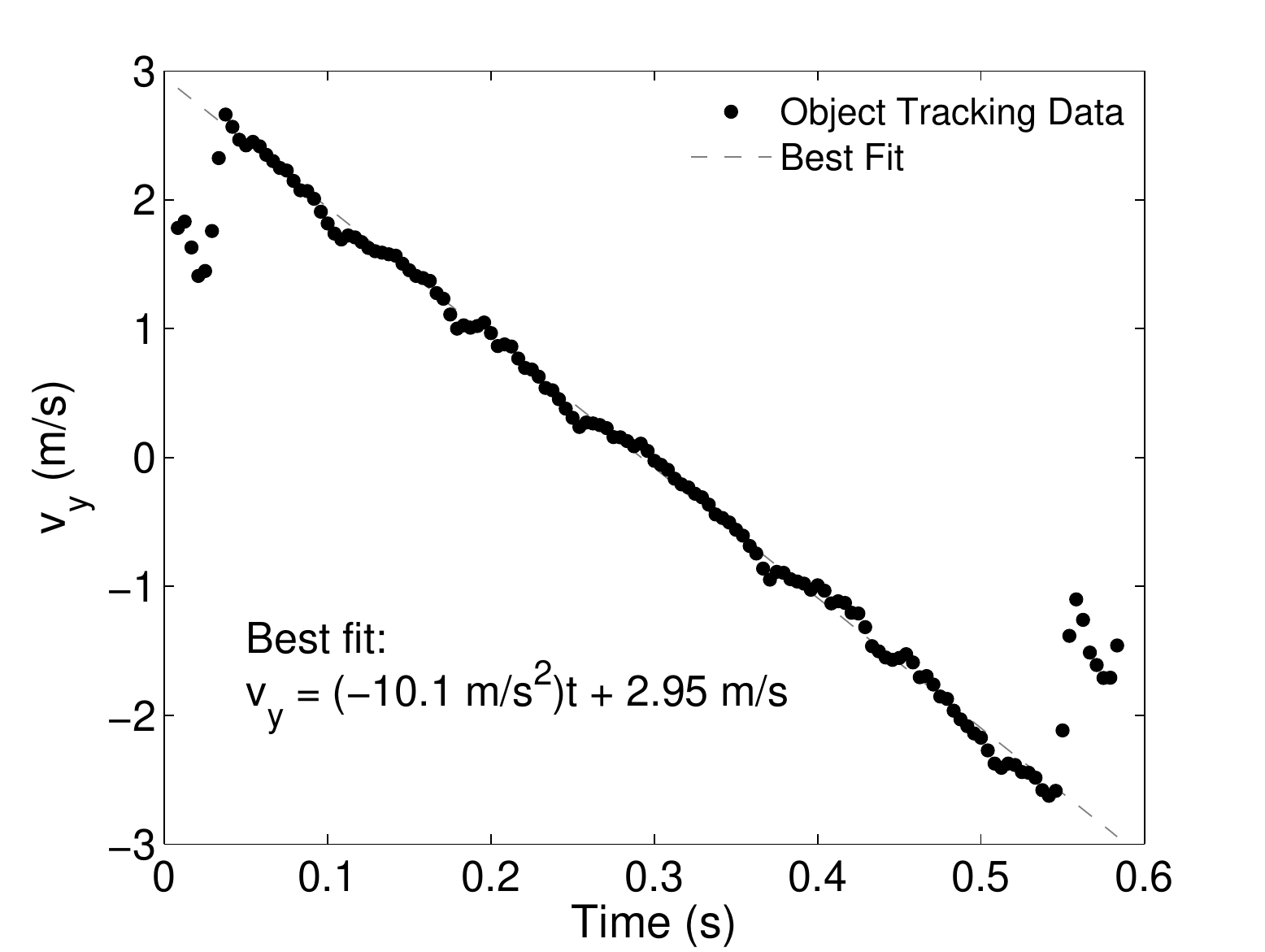}
    \caption{The inferred vertical velocity versus time of the center of mass of the blue painters tape shown in Fig.~\ref{fig:tape}. Also shown is a best fit from a linear regression on the data which gives a slope of $-10.1$~m/s$^2$.}
    \label{fig:vy}
\end{figure}

A plot of the inferred vertical velocity versus time is shown in Fig.~\ref{fig:vy}. The linear nature of the $v_y$ versus time plot confirms the essential usefulness of these color-oriented methods for tracking objects. The best fit implies a gravitational constant of $-10.1$~m/s$^2$ which is reasonably close to reality. The fit excludes the points at the beginning and the end because the tape is moving into the region where the pixels are being tracked which biases the center pixel from the center of mass of the object.

We have not necessarily proven that ``image template" methods will generally fail on this example but we hope that the basic ideas outlined in this paper can be used to improve object tracking methods and help make automatic object tracking more widespread and robust. For the wider need to make physics labs more focused on developing laboratory skills (e.g. \cite{Etkina2015,Ansell_Selen2016,Holmes_Wieman2018}), setting up and recording a direct measurement video can be an interesting and engaging way to get students to think about experimental design. Likewise, improved video analysis tools can help connect lab instruction to ``computational thinking" and simple physical simulations as argued in Orban \& Teeling-Smith \cite{OrbanTS}.

\section{Drawbacks / Practicalities}

Because the methods described here make no assumption about where the object is located, each pixel in each frame within a user defined rectangular region of interest needs to be scanned and calculations performed on the color data, which, for high resolution videos, is a time- and memory-consuming task. However this can be done in quasi-real time on typical computers (e.g. Chromebooks) if the video resolution is coarsened. We developed an online video converter at \url{http://go.osu.edu/convert2gif} where students can upload mp4 or mov format videos and download a coarse resolution gif file.

The STEMcoding Object Tracker works best on the Chrome and Edge browsers. Other browsers have difficulty freeing up the RAM after scanning through a frame. The STEMcoding Object Tracker has been tested and used successfully on Windows, MacOS, iPad, Chromebooks, and Linux.

\section{Conclusions}

Automated object tracking in direct measurement videos is a difficult problem from a computational point of view. The problem can be made simpler by using a solid colored background and a brightly colored object in the video recording. Computationally, it is also much simpler to track colors than it is to track a ``template image" and we outline two methods for identifying pixels that are part of the object from the pixel color alone. We demonstrate the usefulness of these methods on the tricky problem of a hollow, blue rotating object as it moves through the air. We implement these algorithms into a program called the STEMcoding Object Tracker which is freely available to use at \url{http://go.osu.edu/objecttracker}.






\newpage

\acknowledgments

We acknowledge support from the 2021 AIP Meggers Award, the Big Idea Seed Grant from the OSU College of Education and Human Ecology, and the Sit Lux grant program at the University of Mount Union. JB was supported by the Department of Defense High Performance Computing intern program. Many thanks to Brian Lane for useful conversations.

\bibliographystyle{unsrt}
\bibliography{main}

\end{document}